\documentclass[%
 aip,
 jmp,%
 amsmath,amssymb,
 reprint,%
]{revtex4-1}

\usepackage{graphicx}
\usepackage[caption=false]{subfig} 
\usepackage{dcolumn}
\usepackage{bm}

\begin{document}

\preprint{AIP/123-QED}

\title[]{Cavity enhanced absorption spectroscopy in the mid-infrared using a supercontinuum source}

\author{Caroline Amiot}
\email[]{caroline.amiot@tut.fi}
\altaffiliation{}
\affiliation{Tampere University of Technology, Optics Laboratory, Physics department, P.O. Box 692, FI-33101 Tampere, Finland}
\affiliation{Institut FEMTO-ST, UMR 6174 CNRS-Universit\'{e} Bourgogne Franche-Comt\'{e}, 15B avenue des Montboucons, 25030 Besan\c{c}on Cedex, France}
\author{Antti Aalto}
\altaffiliation{}
\affiliation{Tampere University of Technology, Optics Laboratory, Physics department, P.O. Box 692, FI-33101 Tampere, Finland}
\author{Piotr Ryczkowski}
\altaffiliation{}
\affiliation{Tampere University of Technology, Optics Laboratory, Physics department, P.O. Box 692, FI-33101 Tampere, Finland}
\author{Juha Toivonen}
\altaffiliation{}
\affiliation{Tampere University of Technology, Optics Laboratory, Physics department, P.O. Box 692, FI-33101 Tampere, Finland}
\author{Go\"{e}ry Genty}
\altaffiliation{}
\affiliation{Tampere University of Technology, Optics Laboratory, Physics department, P.O. Box 692, FI-33101 Tampere, Finland}

%

\date{\today}

\begin{abstract}
We demonstrate incoherent broadband cavity enhanced absorption spectroscopy in the mid-infrared wavelength range from 3000 to 3450 nm using an all-fiber based supercontinuum source. Multi-components gas detection is performed and concentrations of acetylene and methane are retrieved with sub-ppm accuracy. A linear response to nominal gas concentrations is observed demonstrating the feasibility of the method for sensing applications. 
%
\end{abstract}

\maketitle



Gas detection and accurate concentration measurements are important in many fields ranging from industrial process to emission control and pollution monitoring. Different spectroscopic methods have been developed to retrieve gas concentrations with very high accuracy including cavity ring down spectroscopy \cite{Keefe1988, Long:16} and its broadband implementation \cite{Thorpe2006}, integrated cavity output spectroscopy \cite{ICOSKeefe}, noise-immune cavity-enhanced optical-heterodyne molecular spectroscopy \cite{Ye1998}, or cavity enhanced absorption spectroscopy (CEAS) \cite{Engeln1998, Fiedler2003}. Each of these methods presents advantages and drawbacks in terms of sensitivity, selectivity, footprint and cost. 

Cavity enhanced absorption spectroscopy is conceptually relatively simple and a robust experimental setup can be implemented from off-the-shelf components. In CEAS, one uses a highly reflective cavity to increase significantly the optical path and thus the interaction length between the light beam and gas molecules, which leads to enhanced sensitivity. However, because of the mirrors highly reflectivity, the light intensity at the cavity output is dramatically reduced such that a detector with high sensitivity is generally required to measure the absorption. CEAS can be selective for a particular gas absorption line if a source with narrow linewidth is used, or it can also perform multi-components detection when a light source with a broad spectrum is employed. 

\begin{figure}[tp]
\centering
\includegraphics[width=\linewidth]{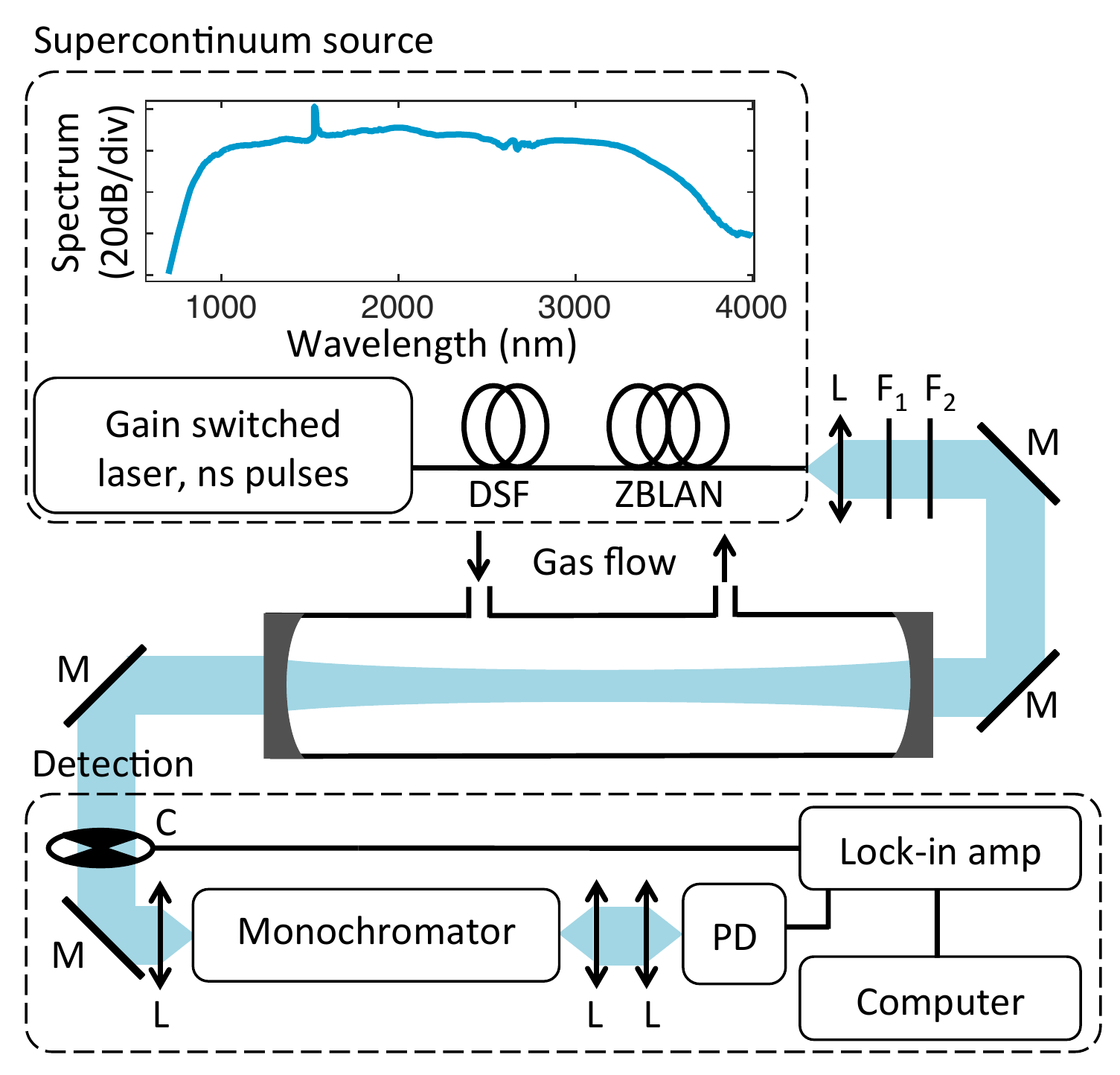}
\caption{Experimental setup. L: lens, F$_{1/2}$: filter, M: mirror, C: mechanical chopper, PD: photodetector.}
\label{fig:setup}
\end{figure}

The recent development of light sources operating in the mid-infrared has recently allowed to extend precise spectroscopic measurements to the molecular fingerprint region where many gases posses strong absorption lines, and indeed several studies have reported measurements from pure gas in the 3-5 microns region \cite{Ye:16, Richter1998, Armstrong:10, Mouawad2016, Chen2001}.  All these recent studies used optical parametric oscillators based on difference-frequency generation or a quantum cascade laser. Whilst some of the recent demonstrations allow for extreme sensitivity, the light source is single specie specific, which may limit the usability.  The development of broadband supercontinuum sources\cite{Kubat:13} on the other hand, has revolutionized many applications ranging from frequency metrology to imaging and spectroscopy. Taking advantage of the high spatial coherence and high brightness of this type of source we demonstrate multi-components gas detection in the mid-infrared over a bandwidth as large as 450~nm using incoherent broadband cavity enhanced absorption spectroscopy. These results are significant not only because they illustrate the potential of incoherent supercontinuum sources for spectroscopy in the mid-IR but also because they represent the largest continuous detection window for gas reported using CEAS method.


An all-fiber supercontinuum source with a spectrum spanning from 900 to 3700~nm was developed using a 1547~nm gain-switched pump laser producing 10~kW peak power sub-nanosecond pulses at a repetition rate of 100~kHz. The pulses are first injected into the anomalous dispersion regime of a 4~m long step-index, silica non-zero dispersion shifted fiber (DSF) with 9~$\mu$m core-diameter. The zero dispersion wavelength (ZDW) of the fiber is at 1510~nm, allowing for efficient noise-seeded modulation instability dynamics which breaks up the long pump pulses into a large number of solitons with short duration \cite{Dudley2006}. The solitons experience the Raman self-frequency shift which expands the spectrum towards the long wavelengths up to 2400~nm, which is the transparency window limit of silica. The output of the DSF is then directly connected to a 7-m step-index fluoride (ZBLAN) fiber with low attenuation ($<0 .1$ dB/m) up to c.a. 4~$\mu$m. The core size of the ZBLAN fiber is comparable to that of the DSF which minimizes the coupling losses. Because a significant fraction of the spectral intensity after the DSF is located in the anomalous dispersion regime of the ZBLAN (ZDW$_{\mathrm{ZBLAN}} $ = 1620~nm) the short solitons undergo additional Raman self-frequency shift extending the supercontinuum spectrum up to 3700~nm with an average output power of 160~mW.  The resulting SC spectrum illustrated in Fig.~1 is essentially spectrally incoherent with large fluctuations from pulse to pulse. Yet, the average spectrum is highly stable, a pre-requisite in the context of spectroscopic measurements.  

The all-fiber SC source was subsequently used to perform cavity-enhanced absorption spectroscopy measurements. The experimental setup is illustrated in Fig.~\ref{fig:setup}. The SC beam is collimated to a 3 mm size using an aspheric lens coated for 3 to 5~$\mu$m range. The beam propagates through a 1-m long confocal cavity constructed from an aluminium pipe and closed by two mirrors with 1~m radius of curvature and high reflection coefficient (R $>$ 99.95) between 3000 and 3450~nm. The SC power corresponding to the cavity spectral bandwidth is 4~mW. The high reflectivity of the mirrors results in a transmission of $2.5 \times 10^{-4}$ and an effective path length of c.a. 300~m. In order to match the supercontinuum source spectrum with the spectral bandwidth of the cavity, the short wavelengths components of the SC are filtered out using two long-pass filters F$_{1}$ and F$_{2}$ with cut-off wavelengths at 2333~nm and 2998~nm, respectively. The cavity has valves in both ends allowing gas circulation with a constant flow during the measurements. Gas samples were diluted from 0.5 ppm of C$_{2}$H$_{2}$ and 0.25 ppm of CH$_{4}$ in nitrogen using mass flow controllers (5850S, Brooks Instrument). Using an uncoated MgF$_{2}$ lens with a focal length of 50~mm, the beam at the cavity output is focused onto the entrance slit of a computer-controlled monochromator (HR550 Horiba) with 300~groove/mm grating and the light intensity is detected with a liquid nitrogen-cooled InAs photodetector (Judson J10D).  Wavelength scanning is performed in steps of 0.5~nm. Because only a small fraction of the 4~mW SC power is transmitted through the highly reflective cavity,  we use lock-in detection (Lock-in amplifier + mechanical choper at 245~kHz) and an integration time of 1~s to improve the signal-to-noise ratio. 


We next perform measurements of different gases using the differential optical absorption spectroscopy (DOAS) method performed in sequential steps \cite{Aalto:15}. The spectral transmission of the cavity is first calibrated by a reference measurement with only N$_2$ flowing through the cavity such that no absorption is present. A second measurement is subsequently performed with the gas under study and the spectral transmission from the cavity is then compared to that of the reference measurement:
 
\begin{equation} \frac{I(\lambda) }{I_0(\lambda) } = \frac {\rho (\lambda)}{\rho (\lambda) + \tau(\lambda)} 
\label{eq:DOAS}
\end{equation}

where $I(\lambda)$ and $I_0(\lambda)$ correspond to the spectral intensities recorded for the gas under study and N$_2$ in the cavity, respectively. The parameter $\rho$ represents the mirror spectral losses and $\tau$ is the small single-pass absorption through the cavity of length $d_{0}$ and depends on the absorption cross section $\sigma(\lambda)_{j}$ and number density $N_{j}$ of the gas species $j$ present in the cavity as: 

\begin{equation} 
\tau(\lambda) =  \sum_{j}^{ } \sigma(\lambda)_j N_j d_0. 
\label{eq:DOAS2}
\end{equation}

Retrieval of the gas species concentration is then achieved by fitting the experimentally measured ratio $I(\lambda)/I_0(\lambda)$ with Eqs.~\eqref{eq:DOAS}-\eqref{eq:DOAS2} using a least-square method. In the fitting procedure, we use the HITRAN 2012 database for absorption line strenghs and apply the appropriate Voigt broadening of the lines at the measurement temperature and pressure. The model also accounts for the finite resolution of the monochromator and slits width as well as for possible background drift caused e.g. by light intensity or cavity coupling fluctuations during the measurement \cite{Aalto:15}.

\begin{figure}[tp]
\centering
\subfloat{%
  \includegraphics[clip,width=1.05\columnwidth]{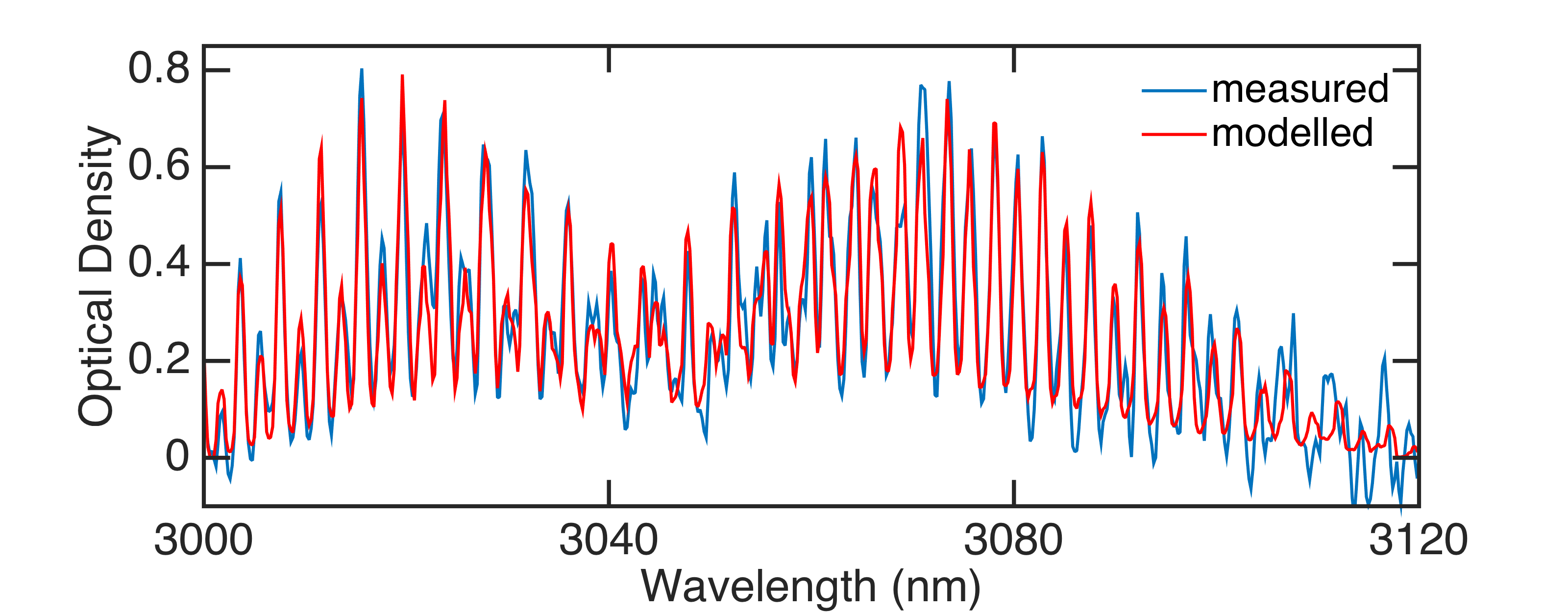}%
}

\subfloat{
  \includegraphics[clip,width=1.05\columnwidth]{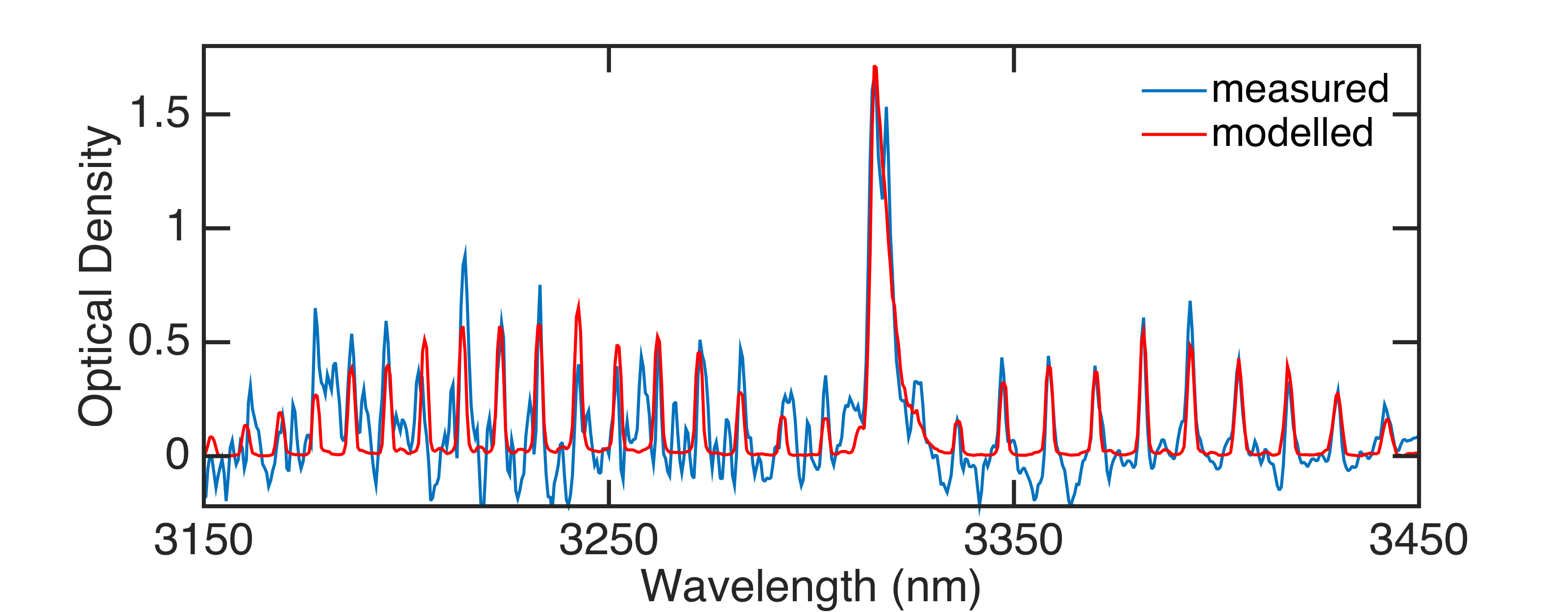}%
}
\caption{Comparison between the modelled (red line) and the measured (blue line) absorption spectra of 5~ppm of acetylene (from 3000 to 3100~nm) and 4~ppm of methane (from 3150 to 3450~nm). Note that negative values in the measured optical density result from the detection noise.}
\label{fig:bothgases}
\end{figure}

One particular feature of broadband DOAS is its ability to retrieve accurately gas concentrations even when the signal-to-noise ratio is relatively low and the recorded absorption spectra noisy. This arises from the fact that spectral fitting is performed over a large number of absorption lines which effectively increases the integration time and thus reduces the noise. 

The experimental setup is first calibrated by measuring the spectrum of a reference gas sample of known concentration, using only the monochromator resolution and mirror losses as free-running parameters. This allows us to determine precisely the spectral dependence of the mirrors losses $\rho(\lambda)$ which may slightly differ from the data provided by the manufacturer.  Figure~\ref{fig:bothgases} shows the comparison between the fitted and measured absorption lines of acetylene (Fig.~\ref{fig:bothgases}a) and methane (Fig.~\ref{fig:bothgases}b) for a nominal concentration of 4~ppm and 5~ppm, respectively, and we can see excellent agreement between the DOAS measurement and the fit. The actual measurement resolution was found to be 1~nm, which is in agreement with the targeted resolution using the monochromator slits of 100~$\mu$m.   

\begin{figure}[hp]
\centering
\subfloat{%
  \includegraphics[clip,width=1.05\columnwidth]{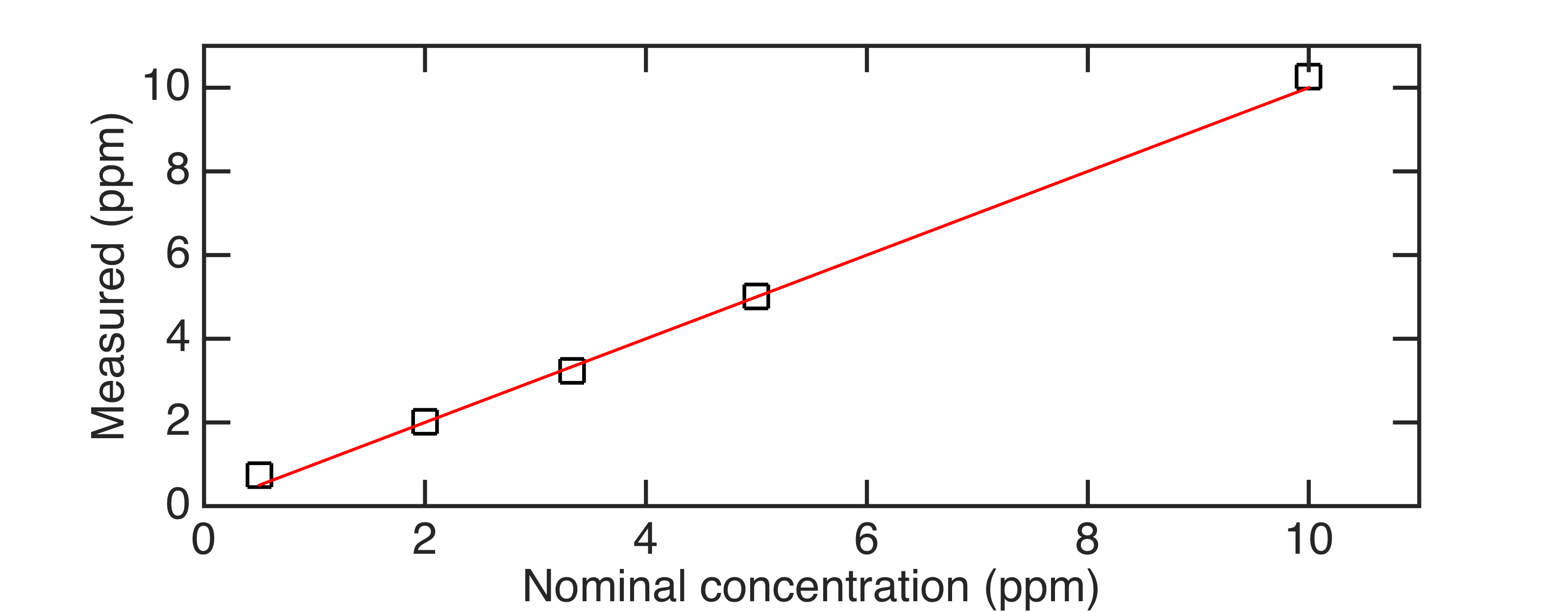}%
}

\subfloat{
  \includegraphics[clip,width=1.05\columnwidth]{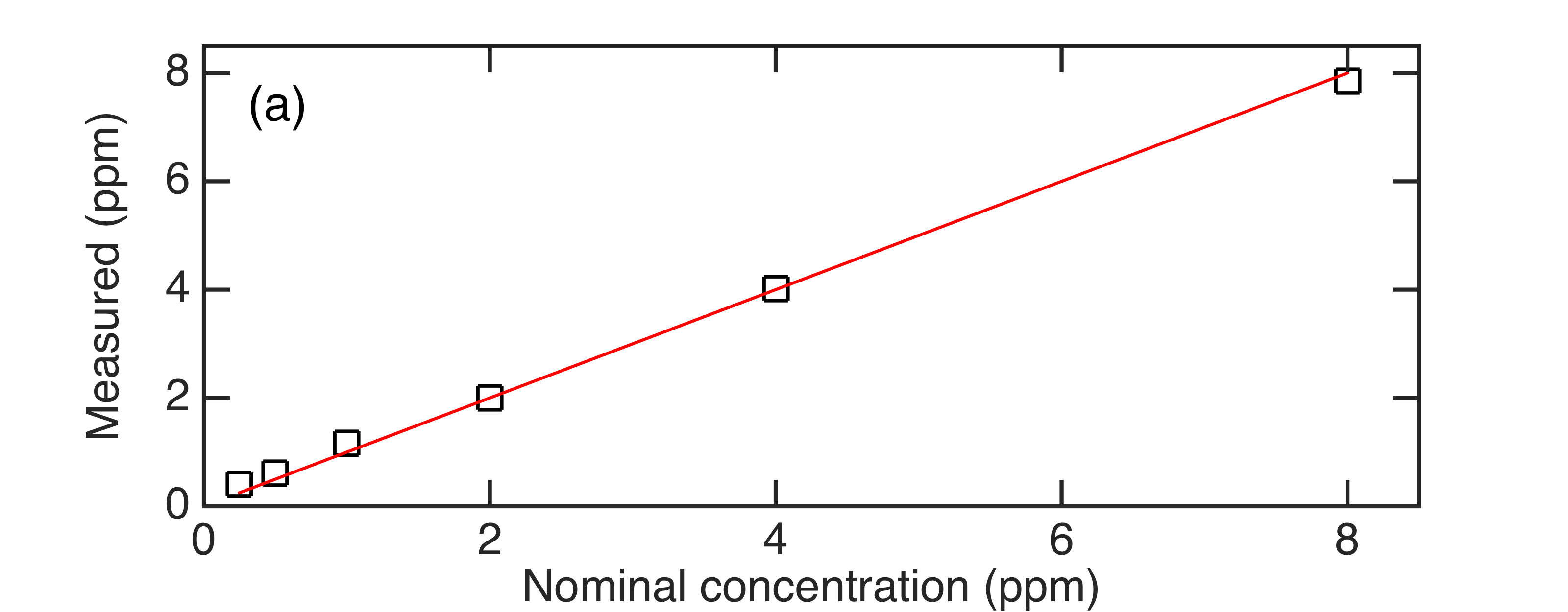}%
}
\caption{Measured vs. nominal concentration for (a) acetylene and (b) methane.}
\label{fig:resolution}
\end{figure}

The calibrated instrument is then used to determine the response of the method by repeating the measurement for different gas concentrations and using the concentration as a free-running parameter in the fitting procedure. Fig.~\ref{fig:resolution} shows the results of the fit against the nominal gas concentrations set using the mass flow controllers. Linear responses are found for both gases demonstrating the feasibility of the method for sensing applications. The smallest concentrations of 0.5 ppm for acetylene and 0.25 ppm for methane are still following the linear response function.  

\begin{figure}[hp]
\centering
\includegraphics[width=1.05\columnwidth]{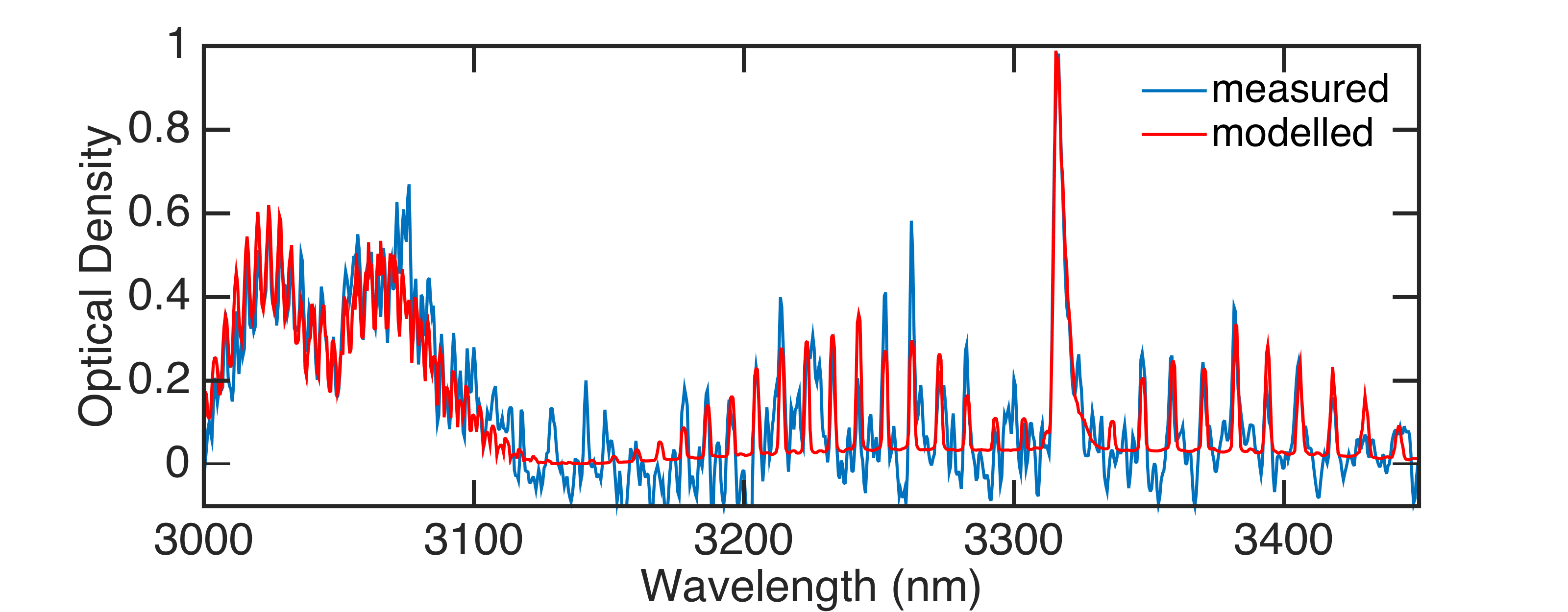}
\caption{Simultaneous multi-component detection of 5~ppm of acetylene (from 3000 to 3150~nm) and 2~ppm of methane (from 3150 to 3450~nm). The blue and red lines illustrate the measured and modelled spectra, respectively. Note that negative values in the measured optical density result from the detection noise.}
\label{fig:multicompo}
\end{figure}

One major advantage of CEAS over other spectroscopic techniques is the possibility to perform detection over a very broad bandwidth. We next exploited this potential to demonstrate multi-component gas analysis over the full bandwidth of the cavity from 3000 to 3450~nm. For this purpose, both methane and acetylene were flown simultaneously through the cavity with nominal concentrations of 5 and 2~ppm, respectively. The results in Fig.~\ref{fig:multicompo} shows very good agreement between the measured and modelled absorption using the nominal concentrations over the full 450~nm bandwidth. For the modelled absorption, only the low-order polynomial was used as used as a free running parameter to account for possible drift during the long measurement time. 

In conclusion, using an all-fiber supercontinuum source we have demonstrated for the first time CEAS in the mid-infrared over a wavelength range as large as 450~nm, allowing to retrieve gas concentrations with sub-ppm accuracy. The sensitivity and measurement speed of the technique could be improved in various ways. The intensity of the light source could be increased e.g. by fusion-splicing the DSF and ZBLAN fibers\cite{Yin2016}, which would reduce the connection losses and allow to extend the spectrum further into the mid-infrared with increased power spectral density. Losses resulting from the non-perfect alignment of the cavity mirrors or from the chromatic aberration of the collimating lens could also be reduced with the use of a parabolic mirror. The measurement speed on the other hand is currently limited to 30~nm/min due to the monochromator scanning and long integration time needed to improve the signal-to-noise ratio.  This could be significantly reduced by using a spectrometer with a detector array \cite{Kim2008}.

This work was supported by the Academy of Finland (Grant 298463) and by the European Union Horizon 2020 research and innovation programme under grant agreement No 722380 SUPUVIR. Caroline Amiot also acknowledges the Doctoral School of Tampere University of Technology.

\bibliography{APL_Amiot}

\begin{thebibliography}{17}%
\makeatletter
\providecommand \@ifxundefined [1]{%
 \@ifx{#1\undefined}
}%
\providecommand \@ifnum [1]{%
 \ifnum #1\expandafter \@firstoftwo
 \else \expandafter \@secondoftwo
 \fi
}%
\providecommand \@ifx [1]{%
 \ifx #1\expandafter \@firstoftwo
 \else \expandafter \@secondoftwo
 \fi
}%
\providecommand \natexlab [1]{#1}%
\providecommand \enquote  [1]{``#1''}%
\providecommand \bibnamefont  [1]{#1}%
\providecommand \bibfnamefont [1]{#1}%
\providecommand \citenamefont [1]{#1}%
\providecommand \href@noop [0]{\@secondoftwo}%
\providecommand \href [0]{\begingroup \@sanitize@url \@href}%
\providecommand \@href[1]{\@@startlink{#1}\@@href}%
\providecommand \@@href[1]{\endgroup#1\@@endlink}%
\providecommand \@sanitize@url [0]{\catcode `\\12\catcode `\$12\catcode
  `\&12\catcode `\#12\catcode `\^12\catcode `\_12\catcode `\%12\relax}%
\providecommand \@@startlink[1]{}%
\providecommand \@@endlink[0]{}%
\providecommand \url  [0]{\begingroup\@sanitize@url \@url }%
\providecommand \@url [1]{\endgroup\@href {#1}{\urlprefix }}%
\providecommand \urlprefix  [0]{URL }%
\providecommand \Eprint [0]{\href }%
\providecommand \doibase [0]{http://dx.doi.org/}%
\providecommand \selectlanguage [0]{\@gobble}%
\providecommand \bibinfo  [0]{\@secondoftwo}%
\providecommand \bibfield  [0]{\@secondoftwo}%
\providecommand \translation [1]{[#1]}%
\providecommand \BibitemOpen [0]{}%
\providecommand \bibitemStop [0]{}%
\providecommand \bibitemNoStop [0]{.\EOS\space}%
\providecommand \EOS [0]{\spacefactor3000\relax}%
\providecommand \BibitemShut  [1]{\csname bibitem#1\endcsname}%
\let\auto@bib@innerbib\@empty
\bibitem [{\citenamefont {O'Keefe}\ and\ \citenamefont
  {Deacon}(1988)}]{Keefe1988}%
  \BibitemOpen
  \bibfield  {author} {\bibinfo {author} {\bibfnamefont {A.}~\bibnamefont
  {O'Keefe}}\ and\ \bibinfo {author} {\bibfnamefont {D.~A.}\ \bibnamefont
  {Deacon}},\ }\bibfield  {title} {\enquote {\bibinfo {title} {Cavity ring-down
  optical spectrometer for absorption measurements using pulsed laser
  sources},}\ }\href@noop {} {\bibfield  {journal} {\bibinfo  {journal} {Review
  of Scientific Instruments}\ }\textbf {\bibinfo {volume} {59}},\ \bibinfo
  {pages} {2544--2551} (\bibinfo {year} {1988})}\BibitemShut {NoStop}%
\bibitem [{\citenamefont {Long}\ \emph {et~al.}(2016)\citenamefont {Long},
  \citenamefont {Fleisher}, \citenamefont {Liu},\ and\ \citenamefont
  {Hodges}}]{Long:16}%
  \BibitemOpen
  \bibfield  {author} {\bibinfo {author} {\bibfnamefont {D.~A.}\ \bibnamefont
  {Long}}, \bibinfo {author} {\bibfnamefont {A.~J.}\ \bibnamefont {Fleisher}},
  \bibinfo {author} {\bibfnamefont {Q.}~\bibnamefont {Liu}}, \ and\ \bibinfo
  {author} {\bibfnamefont {J.~T.}\ \bibnamefont {Hodges}},\ }\bibfield  {title}
  {\enquote {\bibinfo {title} {Ultra-sensitive cavity ring-down spectroscopy in
  the mid-infrared spectral region},}\ }\href {\doibase 10.1364/OL.41.001612}
  {\bibfield  {journal} {\bibinfo  {journal} {Opt. Lett.}\ }\textbf {\bibinfo
  {volume} {41}},\ \bibinfo {pages} {1612--1615} (\bibinfo {year}
  {2016})}\BibitemShut {NoStop}%
\bibitem [{\citenamefont {Thorpe}\ \emph {et~al.}(2006)\citenamefont {Thorpe},
  \citenamefont {Moll}, \citenamefont {Jones}, \citenamefont {Safdi},\ and\
  \citenamefont {Ye}}]{Thorpe2006}%
  \BibitemOpen
  \bibfield  {author} {\bibinfo {author} {\bibfnamefont {M.~J.}\ \bibnamefont
  {Thorpe}}, \bibinfo {author} {\bibfnamefont {K.~D.}\ \bibnamefont {Moll}},
  \bibinfo {author} {\bibfnamefont {R.~J.}\ \bibnamefont {Jones}}, \bibinfo
  {author} {\bibfnamefont {B.}~\bibnamefont {Safdi}}, \ and\ \bibinfo {author}
  {\bibfnamefont {J.}~\bibnamefont {Ye}},\ }\bibfield  {title} {\enquote
  {\bibinfo {title} {Broadband cavity ringdown spectroscopy for sensitive and
  rapid molecular detection},}\ }\href {\doibase 10.1126/science.1123921}
  {\bibfield  {journal} {\bibinfo  {journal} {Science}\ }\textbf {\bibinfo
  {volume} {311}},\ \bibinfo {pages} {1595--1599} (\bibinfo {year}
  {2006})}\BibitemShut {NoStop}%
\bibitem [{\citenamefont {O'Keefe}(1998)}]{ICOSKeefe}%
  \BibitemOpen
  \bibfield  {author} {\bibinfo {author} {\bibfnamefont {A.}~\bibnamefont
  {O'Keefe}},\ }\bibfield  {title} {\enquote {\bibinfo {title} {Integrated
  cavity output analysis of ultra-weak absorption},}\ }\href@noop {} {\bibfield
   {journal} {\bibinfo  {journal} {Chemical Physics Letters}\ }\textbf
  {\bibinfo {volume} {293}},\ \bibinfo {pages} {331--336} (\bibinfo {year}
  {1998})}\BibitemShut {NoStop}%
\bibitem [{\citenamefont {Ye}, \citenamefont {Ma},\ and\ \citenamefont
  {Hall}(1998)}]{Ye1998}%
  \BibitemOpen
  \bibfield  {author} {\bibinfo {author} {\bibfnamefont {J.}~\bibnamefont
  {Ye}}, \bibinfo {author} {\bibfnamefont {L.-S.}\ \bibnamefont {Ma}}, \ and\
  \bibinfo {author} {\bibfnamefont {J.~L.}\ \bibnamefont {Hall}},\ }\bibfield
  {title} {\enquote {\bibinfo {title} {Ultrasensitive detections in atomic and
  molecular physics: demonstration in molecular overtone spectroscopy},}\
  }\href {\doibase 10.1364/JOSAB.15.000006} {\bibfield  {journal} {\bibinfo
  {journal} {J. Opt. Soc. Am. B}\ }\textbf {\bibinfo {volume} {15}},\ \bibinfo
  {pages} {6--15} (\bibinfo {year} {1998})}\BibitemShut {NoStop}%
\bibitem [{\citenamefont {Engeln}\ \emph {et~al.}(1998)\citenamefont {Engeln},
  \citenamefont {Berden}, \citenamefont {Peeters},\ and\ \citenamefont
  {Meijer}}]{Engeln1998}%
  \BibitemOpen
  \bibfield  {author} {\bibinfo {author} {\bibfnamefont {R.}~\bibnamefont
  {Engeln}}, \bibinfo {author} {\bibfnamefont {G.}~\bibnamefont {Berden}},
  \bibinfo {author} {\bibfnamefont {R.}~\bibnamefont {Peeters}}, \ and\
  \bibinfo {author} {\bibfnamefont {G.}~\bibnamefont {Meijer}},\ }\bibfield
  {title} {\enquote {\bibinfo {title} {Cavity enhanced absorption and cavity
  enhanced magnetic rotation spectroscopy},}\ }\href@noop {} {\bibfield
  {journal} {\bibinfo  {journal} {Review of Scientific Instruments}\ }\textbf
  {\bibinfo {volume} {69}},\ \bibinfo {pages} {3763--3769} (\bibinfo {year}
  {1998})}\BibitemShut {NoStop}%
\bibitem [{\citenamefont {Fiedler}, \citenamefont {Hese},\ and\ \citenamefont
  {Ruth}(2003)}]{Fiedler2003}%
  \BibitemOpen
  \bibfield  {author} {\bibinfo {author} {\bibfnamefont {S.~E.}\ \bibnamefont
  {Fiedler}}, \bibinfo {author} {\bibfnamefont {A.}~\bibnamefont {Hese}}, \
  and\ \bibinfo {author} {\bibfnamefont {A.~A.}\ \bibnamefont {Ruth}},\
  }\bibfield  {title} {\enquote {\bibinfo {title} {Incoherent broad-band
  cavity-enhanced absorption spectroscopy},}\ }\href@noop {} {\bibfield
  {journal} {\bibinfo  {journal} {Chemical physics letters}\ }\textbf {\bibinfo
  {volume} {371}},\ \bibinfo {pages} {284--294} (\bibinfo {year}
  {2003})}\BibitemShut {NoStop}%
\bibitem [{\citenamefont {Ye}\ \emph {et~al.}(2016)\citenamefont {Ye},
  \citenamefont {Li}, \citenamefont {Zheng}, \citenamefont {Sanchez},
  \citenamefont {Gluszek}, \citenamefont {Hudzikowski}, \citenamefont {Dong},
  \citenamefont {Griffin},\ and\ \citenamefont {Tittel}}]{Ye:16}%
  \BibitemOpen
  \bibfield  {author} {\bibinfo {author} {\bibfnamefont {W.}~\bibnamefont
  {Ye}}, \bibinfo {author} {\bibfnamefont {C.}~\bibnamefont {Li}}, \bibinfo
  {author} {\bibfnamefont {C.}~\bibnamefont {Zheng}}, \bibinfo {author}
  {\bibfnamefont {N.~P.}\ \bibnamefont {Sanchez}}, \bibinfo {author}
  {\bibfnamefont {A.~K.}\ \bibnamefont {Gluszek}}, \bibinfo {author}
  {\bibfnamefont {A.~J.}\ \bibnamefont {Hudzikowski}}, \bibinfo {author}
  {\bibfnamefont {L.}~\bibnamefont {Dong}}, \bibinfo {author} {\bibfnamefont
  {R.~J.}\ \bibnamefont {Griffin}}, \ and\ \bibinfo {author} {\bibfnamefont
  {F.~K.}\ \bibnamefont {Tittel}},\ }\bibfield  {title} {\enquote {\bibinfo
  {title} {Mid-infrared dual-gas sensor for simultaneous detection of methane
  and ethane using a single continuous-wave interband cascade laser},}\ }\href
  {\doibase 10.1364/OE.24.016973} {\bibfield  {journal} {\bibinfo  {journal}
  {Opt. Express}\ }\textbf {\bibinfo {volume} {24}},\ \bibinfo {pages}
  {16973--16985} (\bibinfo {year} {2016})}\BibitemShut {NoStop}%
\bibitem [{\citenamefont {Richter}\ \emph {et~al.}(1998)\citenamefont
  {Richter}, \citenamefont {Lancaster}, \citenamefont {Curl}, \citenamefont
  {Neu},\ and\ \citenamefont {Tittel}}]{Richter1998}%
  \BibitemOpen
  \bibfield  {author} {\bibinfo {author} {\bibfnamefont {D.}~\bibnamefont
  {Richter}}, \bibinfo {author} {\bibfnamefont {D.}~\bibnamefont {Lancaster}},
  \bibinfo {author} {\bibfnamefont {R.}~\bibnamefont {Curl}}, \bibinfo {author}
  {\bibfnamefont {W.}~\bibnamefont {Neu}}, \ and\ \bibinfo {author}
  {\bibfnamefont {F.}~\bibnamefont {Tittel}},\ }\bibfield  {title} {\enquote
  {\bibinfo {title} {Compact mid-infrared trace gas sensor based on
  difference-frequency generation of two diode lasers in periodically poled
  linbo3},}\ }\href {\doibase 10.1007/s003400050514} {\bibfield  {journal}
  {\bibinfo  {journal} {Applied Physics B}\ }\textbf {\bibinfo {volume} {67}},\
  \bibinfo {pages} {347--350} (\bibinfo {year} {1998})}\BibitemShut {NoStop}%
\bibitem [{\citenamefont {Armstrong}\ \emph {et~al.}(2010)\citenamefont
  {Armstrong}, \citenamefont {Johnstone}, \citenamefont {Duffin}, \citenamefont
  {Lengden}, \citenamefont {Chakraborty},\ and\ \citenamefont
  {Ruxton}}]{Armstrong:10}%
  \BibitemOpen
  \bibfield  {author} {\bibinfo {author} {\bibfnamefont {I.}~\bibnamefont
  {Armstrong}}, \bibinfo {author} {\bibfnamefont {W.}~\bibnamefont
  {Johnstone}}, \bibinfo {author} {\bibfnamefont {K.}~\bibnamefont {Duffin}},
  \bibinfo {author} {\bibfnamefont {M.}~\bibnamefont {Lengden}}, \bibinfo
  {author} {\bibfnamefont {A.~L.}\ \bibnamefont {Chakraborty}}, \ and\ \bibinfo
  {author} {\bibfnamefont {K.}~\bibnamefont {Ruxton}},\ }\bibfield  {title}
  {\enquote {\bibinfo {title} {Detection of ch4 in the mid-ir using difference
  frequency generation with tunable diode laser spectroscopy},}\ }\href
  {http://jlt.osa.org/abstract.cfm?URI=jlt-28-10-1435} {\bibfield  {journal}
  {\bibinfo  {journal} {J. Lightwave Technol.}\ }\textbf {\bibinfo {volume}
  {28}},\ \bibinfo {pages} {1435--1442} (\bibinfo {year} {2010})}\BibitemShut
  {NoStop}%
\bibitem [{\citenamefont {Mouawad}\ \emph {et~al.}(2016)\citenamefont
  {Mouawad}, \citenamefont {B{\'e}jot}, \citenamefont {Billard}, \citenamefont
  {Mathey}, \citenamefont {Kibler}, \citenamefont {D{\'e}s{\'e}v{\'e}davy},
  \citenamefont {Gadret}, \citenamefont {Jules}, \citenamefont {Faucher},\ and\
  \citenamefont {Smektala}}]{Mouawad2016}%
  \BibitemOpen
  \bibfield  {author} {\bibinfo {author} {\bibfnamefont {O.}~\bibnamefont
  {Mouawad}}, \bibinfo {author} {\bibfnamefont {P.}~\bibnamefont {B{\'e}jot}},
  \bibinfo {author} {\bibfnamefont {F.}~\bibnamefont {Billard}}, \bibinfo
  {author} {\bibfnamefont {P.}~\bibnamefont {Mathey}}, \bibinfo {author}
  {\bibfnamefont {B.}~\bibnamefont {Kibler}}, \bibinfo {author} {\bibfnamefont
  {F.}~\bibnamefont {D{\'e}s{\'e}v{\'e}davy}}, \bibinfo {author} {\bibfnamefont
  {G.}~\bibnamefont {Gadret}}, \bibinfo {author} {\bibfnamefont {J.-C.}\
  \bibnamefont {Jules}}, \bibinfo {author} {\bibfnamefont {O.}~\bibnamefont
  {Faucher}}, \ and\ \bibinfo {author} {\bibfnamefont {F.}~\bibnamefont
  {Smektala}},\ }\bibfield  {title} {\enquote {\bibinfo {title}
  {Filament-induced visible-to-mid-ir supercontinuum in a znse crystal: Towards
  multi-octave supercontinuum absorption spectroscopy},}\ }\href {\doibase
  http://dx.doi.org/10.1016/j.optmat.2016.08.009} {\bibfield  {journal}
  {\bibinfo  {journal} {Optical Materials}\ }\textbf {\bibinfo {volume} {60}},\
  \bibinfo {pages} {355 -- 358} (\bibinfo {year} {2016})}\BibitemShut {NoStop}%
\bibitem [{\citenamefont {Chen}\ \emph {et~al.}(2001)\citenamefont {Chen},
  \citenamefont {Mouret}, \citenamefont {Boucher},\ and\ \citenamefont
  {Tittel}}]{Chen2001}%
  \BibitemOpen
  \bibfield  {author} {\bibinfo {author} {\bibfnamefont {W.}~\bibnamefont
  {Chen}}, \bibinfo {author} {\bibfnamefont {G.}~\bibnamefont {Mouret}},
  \bibinfo {author} {\bibfnamefont {D.}~\bibnamefont {Boucher}}, \ and\
  \bibinfo {author} {\bibfnamefont {F.}~\bibnamefont {Tittel}},\ }\bibfield
  {title} {\enquote {\bibinfo {title} {Mid-infrared trace gas detection using
  continuous-wave difference frequency generation in periodically poled
  rbtioaso4},}\ }\href {\doibase 10.1007/s003400100561} {\bibfield  {journal}
  {\bibinfo  {journal} {Applied Physics B}\ }\textbf {\bibinfo {volume} {72}},\
  \bibinfo {pages} {873--876} (\bibinfo {year} {2001})}\BibitemShut {NoStop}%
\bibitem [{\citenamefont {Kubat}\ \emph {et~al.}(2013)\citenamefont {Kubat},
  \citenamefont {Agger}, \citenamefont {Moselund},\ and\ \citenamefont
  {Bang}}]{Kubat:13}%
  \BibitemOpen
  \bibfield  {author} {\bibinfo {author} {\bibfnamefont {I.}~\bibnamefont
  {Kubat}}, \bibinfo {author} {\bibfnamefont {C.~S.}\ \bibnamefont {Agger}},
  \bibinfo {author} {\bibfnamefont {P.~M.}\ \bibnamefont {Moselund}}, \ and\
  \bibinfo {author} {\bibfnamefont {O.}~\bibnamefont {Bang}},\ }\bibfield
  {title} {\enquote {\bibinfo {title} {Mid-infrared supercontinuum generation
  to 4.5 $\mu$m in uniform and tapered zblan step-index fibers by direct
  pumping at 1064 or 1550 nm},}\ }\href {\doibase 10.1364/JOSAB.30.002743}
  {\bibfield  {journal} {\bibinfo  {journal} {J. Opt. Soc. Am. B}\ }\textbf
  {\bibinfo {volume} {30}},\ \bibinfo {pages} {2743--2757} (\bibinfo {year}
  {2013})}\BibitemShut {NoStop}%
\bibitem [{\citenamefont {Dudley}, \citenamefont {Genty},\ and\ \citenamefont
  {Coen}(2006)}]{Dudley2006}%
  \BibitemOpen
  \bibfield  {author} {\bibinfo {author} {\bibfnamefont {J.~M.}\ \bibnamefont
  {Dudley}}, \bibinfo {author} {\bibfnamefont {G.}~\bibnamefont {Genty}}, \
  and\ \bibinfo {author} {\bibfnamefont {S.}~\bibnamefont {Coen}},\ }\bibfield
  {title} {\enquote {\bibinfo {title} {Supercontinuum generation in photonic
  crystal fiber},}\ }\href@noop {} {\bibfield  {journal} {\bibinfo  {journal}
  {Rev. Mod. Phys.}\ }\textbf {\bibinfo {volume} {78}},\ \bibinfo {pages}
  {1135--1184} (\bibinfo {year} {2006})}\BibitemShut {NoStop}%
\bibitem [{\citenamefont {Aalto}\ \emph {et~al.}(2015)\citenamefont {Aalto},
  \citenamefont {Genty}, \citenamefont {Laurila},\ and\ \citenamefont
  {Toivonen}}]{Aalto:15}%
  \BibitemOpen
  \bibfield  {author} {\bibinfo {author} {\bibfnamefont {A.}~\bibnamefont
  {Aalto}}, \bibinfo {author} {\bibfnamefont {G.}~\bibnamefont {Genty}},
  \bibinfo {author} {\bibfnamefont {T.}~\bibnamefont {Laurila}}, \ and\
  \bibinfo {author} {\bibfnamefont {J.}~\bibnamefont {Toivonen}},\ }\bibfield
  {title} {\enquote {\bibinfo {title} {Incoherent broadband cavity enhanced
  absorption spectroscopy using supercontinuum and superluminescent diode
  sources},}\ }\href {\doibase 10.1364/OE.23.025225} {\bibfield  {journal}
  {\bibinfo  {journal} {Opt. Express}\ }\textbf {\bibinfo {volume} {23}},\
  \bibinfo {pages} {25225--25234} (\bibinfo {year} {2015})}\BibitemShut
  {NoStop}%
\bibitem [{\citenamefont {Yin}\ \emph {et~al.}(2016)\citenamefont {Yin},
  \citenamefont {Zhang}, \citenamefont {Yao}, \citenamefont {Yang},
  \citenamefont {Chen},\ and\ \citenamefont {Hou}}]{Yin2016}%
  \BibitemOpen
  \bibfield  {author} {\bibinfo {author} {\bibfnamefont {K.}~\bibnamefont
  {Yin}}, \bibinfo {author} {\bibfnamefont {B.}~\bibnamefont {Zhang}}, \bibinfo
  {author} {\bibfnamefont {J.}~\bibnamefont {Yao}}, \bibinfo {author}
  {\bibfnamefont {L.}~\bibnamefont {Yang}}, \bibinfo {author} {\bibfnamefont
  {S.}~\bibnamefont {Chen}}, \ and\ \bibinfo {author} {\bibfnamefont
  {J.}~\bibnamefont {Hou}},\ }\bibfield  {title} {\enquote {\bibinfo {title}
  {Highly stable, monolithic, single-mode mid-infrared supercontinuum source
  based on low-loss fusion spliced silica and fluoride fibers},}\ }\href
  {\doibase 10.1364/OL.41.000946} {\bibfield  {journal} {\bibinfo  {journal}
  {Opt. Lett.}\ }\textbf {\bibinfo {volume} {41}},\ \bibinfo {pages} {946--949}
  (\bibinfo {year} {2016})}\BibitemShut {NoStop}%
\bibitem [{\citenamefont {Kim}\ \emph {et~al.}(2008)\citenamefont {Kim},
  \citenamefont {Plis}, \citenamefont {Rodriguez}, \citenamefont {Bishop},
  \citenamefont {Sharma}, \citenamefont {Dawson}, \citenamefont {Krishna},
  \citenamefont {Bundas}, \citenamefont {Cook}, \citenamefont {Burrows} \emph
  {et~al.}}]{Kim2008}%
  \BibitemOpen
  \bibfield  {author} {\bibinfo {author} {\bibfnamefont {H.}~\bibnamefont
  {Kim}}, \bibinfo {author} {\bibfnamefont {E.}~\bibnamefont {Plis}}, \bibinfo
  {author} {\bibfnamefont {J.}~\bibnamefont {Rodriguez}}, \bibinfo {author}
  {\bibfnamefont {G.}~\bibnamefont {Bishop}}, \bibinfo {author} {\bibfnamefont
  {Y.}~\bibnamefont {Sharma}}, \bibinfo {author} {\bibfnamefont
  {L.}~\bibnamefont {Dawson}}, \bibinfo {author} {\bibfnamefont
  {S.}~\bibnamefont {Krishna}}, \bibinfo {author} {\bibfnamefont
  {J.}~\bibnamefont {Bundas}}, \bibinfo {author} {\bibfnamefont
  {R.}~\bibnamefont {Cook}}, \bibinfo {author} {\bibfnamefont {D.}~\bibnamefont
  {Burrows}},  \emph {et~al.},\ }\bibfield  {title} {\enquote {\bibinfo {title}
  {Mid-ir focal plane array based on type-ii inas/gasb strain layer
  superlattice detector with nbn design},}\ }\href@noop {} {\bibfield
  {journal} {\bibinfo  {journal} {Applied Physics Letters}\ }\textbf {\bibinfo
  {volume} {92}},\ \bibinfo {pages} {183502} (\bibinfo {year}
  {2008})}\BibitemShut {NoStop}%
\end{thebibliography}%

\end{document}